\documentclass[aps,preprint,preprintnumbers,amsmath,amssymb,floatfix]{revtex4}
\usepackage{epsfig}
\usepackage{graphicx}
\usepackage{bm}
\usepackage{amssymb}
\usepackage{dcolumn}% Align table columns on decimal point
\usepackage{bm}% bold math
\begin{document}
\draft

\title
{Use of vector polarizability to manipulate alkali-metal atoms}
\author{D. Cho\footnote{e-mail address:{\tt cho@korea.ac.kr}}}
\affiliation{Department of Physics, Korea University, Seoul 02841, Korea}

\date{\today}

\begin{abstract}
We review a few ideas and experiments that our laboratory at Korea University has proposed 
and carried out to use vector polarizability $\beta$ to manipulate alkali-metal atoms. 
$\beta$ comes from spin-orbit coupling, and it produces an ac Stark shift 
that resembles a Zeeman shift. 
When a circularly polarized laser field is properly detuned between the $D_1$ and $D_2$ transitions, 
an ac Stark shift of a ground-state atom takes the form of a pure Zeeman shift. 
We call it the ``analogous Zeeman effect", and experimentally demonstrated 
an optical Stern-Gerlach effect and an optical trap that behaves exactly like a  magnetic trap.
By tuning polarization of a trapping beam, and thereby controlling a shift proportional to $\beta$, 
we demonstrated elimination of an inhomogeneous broadening of a ground hyperfine transition
in an optical trap. We call it ``magic polarization".
We also showed significant narrowing of a Raman sideband transition at a specific well depth.
A Raman sideband in an optical trap is broadened owing to anharmonicity of the trap potential, 
and the broadening can be eliminated by a $\beta$-induced differential ac Stark shift at 
what we call a ``magic well depth".
Finally, we proposed and experimentally demonstrated a cooling scheme that incorporated
the idea of velocity-selective coherent population trapping to Raman sideband cooling
to enhance cooling efficiency of the latter outside of the Lamb-Dicke regime. 
We call it ``motion-selective coherent population trapping", 
and $\beta$ is responsible for the selectivity.
We include as Supplementary Material a program file that calculates both scalar and vector polarizabilities of a given alkali-metal atom when the wavelength of an applied field is specified.
It also calculates depth of a potential well and photon-scattering rate of a trapped atom 
in a specific ground state
when power, minimum spot size, and polarization of a trap beam are given.
\end{abstract}

\maketitle

\section{INTRODUCTION}
While a near-resonant laser field applies radiation pressure on an atom, 
an off-resonant one can produce an energy shift by the ac Stark effect. 
The former plays an important role in slowing and cooling atoms at an ambient temperature, and the latter, coupled with an intensity gradient, produces an electric dipole force to trap precooled atoms.
When the laser field is far detuned from allowed transitions, 
an optical dipole trap provides an almost conservative potential well,
which is ideal for precision spectroscopy or quantum manipulation of trapped atoms.
In addition, the ac Stark shift of an alkali-metal atom in an elliptically polarized laser field 
depends on the magnetic quantum number of the atom and degree of circularity of the field.
This dependence gives us a versatile tool to manipulate an alkali-metal atom in an optical trap.
In this paper, I will review a few ideas and experiments 
that our group at Korea University has proposed and performed to use this useful tool.

We first introduce spherical tensor formalism of the ac Stark shift
to quantify its polarization dependence in terms of a vector polarizability.
The theory section is followed by one on the analogous Zeeman effect 
\cite{analogous Zeeman theory, analogous Zeeman exp}, 
where we show that when an applied field is circularly polarized and 
detuned properly between the $D_1$ and $D_2$ transitions, 
an ac Stark shift takes the form of a pure Zeeman shift. 
The intensity gradient of such a field produces optical Stern-Gerlach effect \cite{optical Stern-Gerlach} or
an optical trap that behaves like a magnetic trap \cite{analogous magnetic trap}.
Next two sections describe ``magic" ideas;
(i) A shift proportional to vector polarizability can be used 
to eliminate an inhomogeneous broadening of a ground hyperfine transition, 
which originates from the difference in scalar polarizabilities of the two hyperfine levels.
The elimination is complete at a particular polarization, and we call it magic polarization
\cite{magic pol theory, magic pol experiment}.
(ii) Stimulated Raman transitions of a trapped atom that involve the change in vibrational quantum number 
show inhomogeneous broadening from anharmonicity of an optical trap. 
It can be eliminated by another broadening from the vector polarizability.
Anharmonicity-induced broadening is independent of well depth, 
and we use the well depth as a tuning parameter to eliminate the broadening.
We call it magic well depth \cite{magic well depth}.
Finally, we describe a cooling method that we call
motion-selective coherent population trapping
\cite{MSCPT theory, MSCPT exp}.
In this method, the ideas of 
Raman sideband cooling and velocity-selective coherent population trapping are combined
to complement each other,
and vector polarizability plays a pivotal role in producing motion-selectivity.

\section{Spherical Tensor formalism of ac Stark shift}
When an atom in the state $|\Psi_0 \rangle$ is irradiated by a laser field with an oscillating electric field
$\vec{E} (t) = \vec{\mathcal{E}} e^{- i \omega t} + {\vec{\mathcal{E}}}^* e^{i \omega t}$, its ac
Stark shift is, from second-order perturbation theory, 
\begin{equation}
	U_{\rm{AC}} (\Psi_0 ) = \sum_{n \neq 0} 
	\frac{\langle \Psi_0  | \vec{d} \cdot \vec{\mathcal{E}}^* | \Psi_n\rangle 
		\langle \Psi_n  | \vec{d} \cdot \vec{\mathcal{E}} | \Psi_0 \rangle }
	{\hbar[\, \omega-(\omega_n - \omega_0)]} -
	\sum_{n \neq 0} \frac{\langle \Psi_0 | \vec{d} \cdot \vec{\mathcal{E}} | \Psi_n \rangle 
		\langle \Psi_n  | \vec{d}\cdot \vec{\mathcal{E}}^* | \Psi_0 \rangle }
	{\hbar[\,\omega+ (\omega_n-\omega_0)]},
	\label{eq: ac Stark 1}
\end{equation}
where $\omega_0$ and $\omega_n$ are the unperturbed energy eigenvalues of 
$| \Psi_0 \rangle$ and $| \Psi_n \rangle$, respectively, and $\vec{d}$ is the electric dipole moment operator.
When $|\Psi_0 \rangle$ is the ground state, the second term is off resonant and it is much smaller than the first in most cases.
$U_{\rm{AC}} (\Psi_0 )$ can be written as 
$ \langle \Psi_0 | \Omega_++ \Omega_-| \Psi_0 \rangle$, where
\begin{eqnarray}
	\label{Omega_plus}
	\Omega_{+}  &=&  
	\sum_{i, \, j}  d_{i}\mathcal{E}_{i}^{\ast}  (\omega-H + \omega_0)^{-1}d_{j} \mathcal{E}_{j}, \\
	\label{Omega_minus}
	\Omega_-  &=&  
	\sum_{i, \, j}  d_{i}\mathcal{E}_{i}  (-\omega -H + \omega_0)^{-1}d_{j} \mathcal{E}_{j}^{\ast}, 
\end{eqnarray}
where $H$ is the unperturbed atomic Hamiltonian, and 
$\mathcal{E}_i$, $\mathcal{E}_j$ and $d_i$, $d_j$ for $i, j = x, y, z$ are the Cartesian components of $\vec{\mathcal{E}}$ and $\vec{d}$, respectively. 
The operators can be rewritten in a spherical tensor form  \cite{Happer}, and explicitly for $\Omega_+$,
\begin{equation}
	\Omega_+ =  \sum_{L=0}^{2} (-1)^{L}
	\sum_{m=-L}^{L} (-1)^{m}
	{\cal{D}}_{m}^{(L)} 
	{\mathcal{F}}_{-m}^{(L)} ,
\end{equation}
where ${\cal{D}}^{(L)}$ and ${\mathcal{F}}^{(L)}$ with $L$ = 0, 1, 2 are the scalar, vector and the second-rank
spherical tensor operators:
\begin{equation}
	{\cal{D}}_{m}^{(L)} = (-1)^{m+1} \sqrt{2L+1} \sum_{\mu=-1}^{1}
	\left( \begin{array}{ccc} 1 & 1 & L \\  \mu & m-\mu & -m  \end{array} \right)
	d_{\mu}(\omega_0 - H_0 + \omega)^{-1} d_{m-\mu}.
\end{equation}
Here $d_{0}=z$ and $d_{\pm 1}=\mp(x\pm iy)/\sqrt{2}$. 
${\mathcal{F}}_{m}^{(L)}$ are similarly defined in terms of $\mathcal{E}_{i}^{\ast}$ and $\mathcal{E}_{j}$;
\begin{equation}
	{\mathcal{F}}_{m}^{(L)} = (-1)^{m+1} \sqrt{2L+1} \sum_{\mu=-1}^{1}
	\left( \begin{array}{ccc} 1 & 1 & L \\  \mu & m-\mu & -m  \end{array} \right)
	\mathcal{E}_{\mu}^\ast \mathcal{E}_{m-\mu},
\end{equation}
where $\mathcal{E}_{0}=\mathcal{E}_z$ 
and $\mathcal{E}_{\pm 1}=\mp(\mathcal{E}_x \pm i \mathcal{E}_y )/\sqrt{2}$.
$\Omega_-$ can be similarly reformulated.

While the scalar term in $U_{\rm{AC}} (\Psi_0 )$ is simply proportional to $|\vec{\mathcal{E}} |^2$,
the vector term is proportional to 
$\langle \Psi_0  | \vec{\sigma} | \Psi_0 \rangle \cdot (\vec{\mathcal{E}} \times \vec{\mathcal{E}}^{\ast})$
leading to dependence of the ac Stark shift on 
magnetic quantum number of the atom and polarization of the field.
Here, ${\cal{D}}^{(1)}$ in the vector term is replaced with the Pauli spin operator 
using the Wigner-Eckart theorem.
The second-rank term is nonvanishing only if the total angular momentum of the state $\Psi_0$ is at least 1. 
Although the ground state of an alkali-metal atom has a nuclear spin in addition to the electron spin 1/2
so that the total angular momentum $F$ can be larger than 1, 
when the detuning of the field $\vec{E} (t)$ from the $D$ transitions is much larger
than the hyperfine splitting, the second-rank term can be neglected.
Under this approximation, the ac Stark shift of the ground hyperfine state 
$|nS_{1/2}, F, m_F \rangle$ of an alkali-metal atom, 
where $m_F$ is the magnetic quantum number, can be written as
\begin{equation}
	U_{\rm{AC}} (F,m_F)= \alpha |\vec{\mathcal{E}}|^2
	 -i \beta \langle S_{1/2},F, m_F|\vec{\sigma}|S_{1/2}, F, m_F\rangle
	 \cdot (\vec{\mathcal{E}} \times \vec{\mathcal{E}}^\ast),
	\label{eq:ac Stark 2}
\end{equation}
where $\alpha$ and $\beta$ are the scalar and the vector polarizabilities, respectively. 
Their expressions can be obtained by comparing Eq. (\ref{eq: ac Stark 1}) with Eq. (\ref{eq:ac Stark 2}):
\begin{eqnarray}
	 \label{eq:alpha} 
	\alpha &=& \frac{1}{6\hbar} \sum_{n^\prime} 
	 {| \langle n^{\prime} P_{1/2} \,||\, d \,||\, nS_{1/2} \rangle|^2}
	   \left[\frac{1}{\omega -(\omega_{n^\prime P_{1/2}}-\omega_{nS_{1/2}})} 
	   -\frac{1}{\omega +(\omega_{n^\prime P_{1/2}}-\omega_{nS_{1/2}})}\right] \\ \nonumber
	&+& \frac{1}{6\hbar} \sum_{n^\prime}
	{| \langle n^{\prime} P_{3/2} \,||\, d \,||\, nS_{1/2} \rangle|^2}
 \left[\frac{1}{\omega -(\omega_{n^\prime P_{3/2}}-\omega_{nS_{1/2}})} 
 -\frac{1}{\omega +(\omega_{n^\prime P_{3/2}}-\omega_{nS_{1/2}})}\right], \\
 \label{eq:beta} 
	\beta &=& \frac{1}{6\hbar} \sum_{n^\prime} 
	{| \langle n^{\prime} P_{1/2} \,||\, d \,||\, nS_{1/2} \rangle|^2}
	\left[\frac{1}{\omega -(\omega_{n^\prime P_{1/2}}-\omega_{nS_{1/2}})} 
	+\frac{1}{\omega +(\omega_{n^\prime P_{1/2}}-\omega_{nS_{1/2}})}\right] \\ \nonumber
	&-& \frac{1}{12\hbar} \sum_{n^\prime}
	{| \langle n^{\prime} P_{3/2} \,||\, d \,||\, nS_{1/2} \rangle|^2}
	\left[\frac{1}{\omega -(\omega_{n^\prime P_{3/2}}-\omega_{nS_{1/2}})} 
	+\frac{1}{\omega +(\omega_{n^\prime P_{3/2}}-\omega_{nS_{1/2}})}\right],
\end{eqnarray}
where $\langle n^{\prime} P_{J} \,||\, d \,||\, nS_{1/2} \rangle $ for $J= 1/2$ and 3/2  is 
the reduced matrix element of $\vec{d}$.
The definition of $\langle n^{\prime} P_{J} \,||\, d \,||\, nS_{1/2} \rangle $ we use is given in Appendix A.

%\newpage
The ac Stark shift in Eq. (\ref{eq:ac Stark 2}) can be written in a form 
which clearly shows its dependence on $m_F$ and polarization of the laser beam:
\begin{equation}
	U_{\rm{AC}} (F,m_F)= (\alpha-\eta\beta g_F m_F)(c\mu_0 I_0/2),
	\label{eq:ac Stark 3}
\end{equation}
where $I_0= 2| \vec{\mathcal{E}}|^2/c\mu_0$ is the intensity of the laser field 
$\vec{E} (t) = \vec{\mathcal{E}} e^{- i \omega t} + {\vec{\mathcal{E}}}^* e^{i \omega t}$,
$g_F$ is the Land\'{e} $g$ factor, and $\eta$ is the degree of circularity:
\begin{equation}
	\eta = i \hat{z} \cdot (\hat{\epsilon} \times \hat{\epsilon}^\ast).
	\label{eq:eta}
\end{equation}
Here $\hat{\epsilon}$ is a Jones vector for the field $\vec{\mathcal{E}}$ 
and $\hat{z}$ is the quantization axis that defines $m_F$.
$\eta$ is 0 for a linearly polarized light, 
and $+1$ and $-1$ for a right and a left circularly polarized light, respectively,
when the light propagates along $\hat{z}$.
We note that the second term has the same form as a Zeeman shift, 
$g_F m_F \mu_B B_0$, of an atom in the $|nS_{1/2}, F, m_F \rangle$ state inside a magnetic field
$B_0\hat{z}$, and we define an effective magnetic field,
\begin{equation}
	\mathcal{B}_{\rm eff}=-\frac{\eta\beta c \mu_0I_0}{2\mu_{\rm B} },
	\label{eq:effective B}
\end{equation} 
where $\mu_{\rm B}$ is the Bohr magneton. 
To understand the origin of this Zeeman-like ac Stark shift,
we write $U_{\rm{AC}} (F,m_F)$ in Eq. (\ref{eq:ac Stark 2}) as an expectation value of 
an electric dipole interaction Hamiltonian,
$- \vec{d}_{\rm ind} (\vec{\mathcal{E}})\cdot \vec{\mathcal{E}}^\ast$, where
\begin{equation}
\vec{d}_{\rm ind} (\vec{\mathcal{E}}) =
	 -\alpha \vec{\mathcal{E}}
	+i \beta (\vec{\sigma}\times \vec{\mathcal{E}} )
	\label{eq:induced dipole}
\end{equation}
is an induced electric dipole moment. 
While the scalar term is from the usual polarization by $\vec{\mathcal{E}}$ and parallel to the field,
the vector term is from perturbation by the spin-orbit coupling and is perpendicular to the field.
The spin-orbit coupling mediates the interaction 
between the magnetic dipole moment of an electron with $\vec{E} (t)$,
while their direct interaction is forbidden by parity ($P$) and the time-reversal ($T$) symmetry. 
We point out that the second term in Eq. (\ref{eq:induced dipole}) is odd under $P$ and even under $T$
like an electric dipole moment, and the equation satisfies $P$ and $T$ symmetry. 
According to this picture, $\beta$ depends on the strength of the spin-orbit coupling. 
This can be shown by rewriting $\alpha$ and $\beta$ in Eqs. (\ref{eq:alpha}, \ref{eq:beta}) as
\begin{eqnarray}
	\label{eq:alpha2}
	\alpha &\approx& \frac{1}{6\hbar} \sum_{n^\prime} 
	{| \langle n^{\prime} P_{1/2} \,||\, d \,||\, nS_{1/2} \rangle|^2}
	\left[\frac{1}{\Delta_1( n^\prime)} 
	+\frac{2}{\Delta_2(n^\prime)} \right], \\ 
	\label{eq:beta2}	
	\beta &\approx& \frac{1}{6\hbar} \sum_{n^\prime} 
	{| \langle n^{\prime} P_{1/2} \,||\, d \,||\, nS_{1/2} \rangle|^2}
	\left[\frac{1}{\Delta_1(n^\prime)} 
	-\frac{1}{\Delta_2(n^\prime)} \right],
\end{eqnarray}
where $\Delta_1( n^\prime) =\omega -(\omega_{n^\prime P_{1/2}}-\omega_{nS_{1/2}})$
and $\Delta_2( n^\prime) =\omega -(\omega_{n^\prime P_{3/2}}-\omega_{nS_{1/2}})$.
Here we use $| \langle n^{\prime} P_{3/2} \,||\, d \,||\, nS_{1/2} \rangle|^2 \approx
2| \langle n^{\prime} P_{1/2} \,||\, d \,||\, nS_{1/2} \rangle|^2$ and neglect the off-resonant terms. 
Without the spin-orbit coupling, there is no fine structure and $\Delta_1( n^\prime) =\Delta_2( n^\prime)$
and $\beta = 0$.
As a consequence, heavy alkali-metal atoms with a strong spin-orbit coupling 
such as Rb and Cs have larger $\beta$ than the lighter atoms.

%\newpage
In an optical trap, 
in addition to the ac Stark shift $U_{\rm AC}$ in Eq. (\ref{eq: ac Stark 1}), which determines 
well depth, the rate of scattering $R_\gamma$ of the trap beam by an atom,
which determines heating and decoherence rate, is also an important parameter. 
Using first-order time-dependent perturbation theory, the scattering rate is given by
\begin{equation}
	R_{\gamma} (\Psi_0 ) \approx \sum_{n \neq 0} 
	\frac{\langle \Psi_0  | \vec{d} \cdot \vec{\mathcal{E}}^* | \Psi_n\rangle 
		\langle \Psi_n  | \vec{d} \cdot \vec{\mathcal{E}} | \Psi_0 \rangle }
	{\hbar^2\left\{\, \omega-(\omega_n - \omega_0) \right\}^2 }\Gamma(n),
	\label{eq: R_gamma 1}
\end{equation}
where $\Gamma(n)=1/\tau(n)$ with $\tau(n)$ being a lifetime of the upper state $|\Psi_n\rangle$,
and the off-resonant terms are neglected.
If we further assume that $\Gamma(n)$ is independent of $n$, 
$R_{\gamma} (\Psi_0 )$ can be written as $ \langle \Psi_0 | \Theta|\Psi_0 \rangle\Gamma$, 
where
\begin{equation}
	\Theta=
	\sum_{i, \, j}  d_{i}\mathcal{E}_{i}^{\ast}  (\omega-H + \omega_0)^{-2}d_{j} \mathcal{E}_{j},	\label{R_operator}
\end{equation}
which is smilar to $\Omega_{\pm}$ in Eqs. (\ref{Omega_plus}, \ref{Omega_minus}).
When $|\Psi_0 \rangle$ is the ground state of an alkali-metal atom 
and a trap beam is red detuned from the $D$ transitions,
only  the contributions from the $D_1$ and $D_2$ transitions to $R_\gamma(\Psi_0)$ are significant
and we may neglect couplings to the higher excited states.
In addition, decay rates of the $nP_{1/2}$ and $nP_{3/2}$ states are approximately the same.
Under these approximations, 
the transformation of $R_\gamma(\Psi_0)$ in Eq. (\ref{eq: R_gamma 1}) 
to $ \langle \Psi_0 | \Theta|\Psi_0 \rangle\Gamma$ is justified. 
Following the spherical tensor formalism that led to Eq. (\ref{eq:ac Stark 3}),
the photon scattering rate of the ground hyperfine state 
$|nS_{1/2}, F, m_F \rangle$ of an alkali-metal atom can be approximated as
\begin{equation}
	R_\gamma (F,m_F)= (a-\eta b g_F m_F)(c\mu_0 I_0/2)\Gamma,
	\label{eq:R_gamma 2}
\end{equation}
where considering only the $D_1$ and $D_2$ couplings, 
\begin{eqnarray}
	\label{eq:a} 
	a &=& \frac{1}{6\hbar^2} \left[
	\frac{| \langle nP_{1/2} \,||\, d \,||\, nS_{1/2} \rangle|^2 }
	{\{\omega -(\omega_{n_{1/2}}-\omega_{nS_{1/2}})\}^2} 
	+\frac{| \langle n P_{3/2} \,||\, d \,||\, nS_{1/2} \rangle|^2 }
	{\{\omega -(\omega_{n P_{3/2}}-\omega_{nS_{1/2}})\}^2} 
	\right],
	\\
	\label{eq:b} 
	b &=& \frac{1}{12\hbar^2}  \left[
	\frac{2 | \langle n P_{1/2} \,||\, d \,||\, nS_{1/2} \rangle|^2 }
	{\{\omega -(\omega_{n P_{1/2}}-\omega_{nS_{1/2}})\}^2} 
	-\frac{ |\langle nP_{3/2} \,||\, d \,||\, nS_{1/2} \rangle|^2 }
	{\{\omega -(\omega_{n P_{3/2}}-\omega_{nS_{1/2}})\}^2} 
	\right].
\end{eqnarray}

%\newpage
In the Supplementary Material, we include a Microsoft Excel file 
that calculates $\alpha, \beta$ and $a, b$ 
of alkali-metal atoms from lithium to cesium when the wavelength of an applied laser beam is given. 
Because we consider a far-detuned trap beam, 
effects of nuclear spin and isotope shift on 
the polarizabilities and the scattering parameters are negligible. 
For a given ground hyperfine state $|nS_{1/2}, F, m_F \rangle$,
the Excel file calculates well depth $U_{\rm AC}(F, m_F)$ and 
photon scattering rate $R_\gamma(F,m_F)$ of an optical trap 
according to Eq. (\ref{eq:ac Stark 3}) and Eq. (\ref{eq:R_gamma 2}), respectively,
when power, minimum spot size, and degree of circularity $\eta$ of the trap beam are given. 
Formulae and conversion factors used for the items in the Excel file are explained in Appendix A. 

%\newpage
\section{Analogous Zeeman effect}
In a typical optical trap, a trap beam is far red detuned from the $D_1$ transition as shown in Fig. 1(a).
Here, we consider a case of tuning the beam between the $D_1$ and $D_2$  transitions as in Fig. 1(b).
In this case, $\Delta_1(n)$ and $\Delta_2(n)$ have opposite signs, 
and we note from Eqs. (\ref{eq:alpha2}, \ref{eq:beta2}) that
$\alpha$ tends to be suppressed while $\beta$ tends to be enhanced. 
Especially when 
\begin{equation}
   \Delta_2(n) = -2\Delta_1(n),
   \label{eq:analogous Zeeman}
\end{equation}
neglecting couplings other than $D_1$ and $D_2$, 
$\alpha$ vanishes and the ac Stark shift becomes
\begin{equation}
	U_{\rm{AC}} (F,m_F)= -\eta\beta g_F m_F  (c\mu_0 I_0/2).
	\label{eq:ac Stark 4}
\end{equation}
It is identical to a Zeeman shift of the $|nS_{1/2}, F, m_F \rangle$ state of an alkali-metal atom
with an effective magnetic field $\mathcal{B}_{\rm eff}$ given in Eq. (\ref{eq:effective B}).
We call it the analogous Zeeman effect \cite{analogous Zeeman theory}, 
which makes a laser beam play the role of a magnetic field
with the field strength proportional to its intensity. 
This effect gives a versatile tool to manipulate alkali-metal atoms. 
However, it is applicable to only heavy alkali-metal atoms such as Rb and Cs 
with large fine structure.
For the lighter atoms, the small fine structure limits detuning. 
Because $R_\gamma$ in Eq. (\ref{eq: R_gamma 1}) is inversely proportional to the detuning squared, excessive scattering of the trap beam becomes a problem.
In Table I, for each alkali-metal atom,
the wavelength $\lambda_0$ that satisfies the condition of Eq. (\ref{eq:analogous Zeeman}) and
$\beta$ at $\lambda_0$ are given.
$\mathcal{B}_{\rm eff}$ and $R_\gamma$ for a state with $g_Fm_F =1$ are also tabulated
when the laser intensity $I_0$ is $1.4 \times 10^9$ W/m$^2$,
corresponding to laser power of 0.5 W and minimum spot size of 15 $\mu$m,
and with right-circular polarization so that $\eta =1$.

 \begin{figure}[h] \centering
	\includegraphics[scale=0.65]{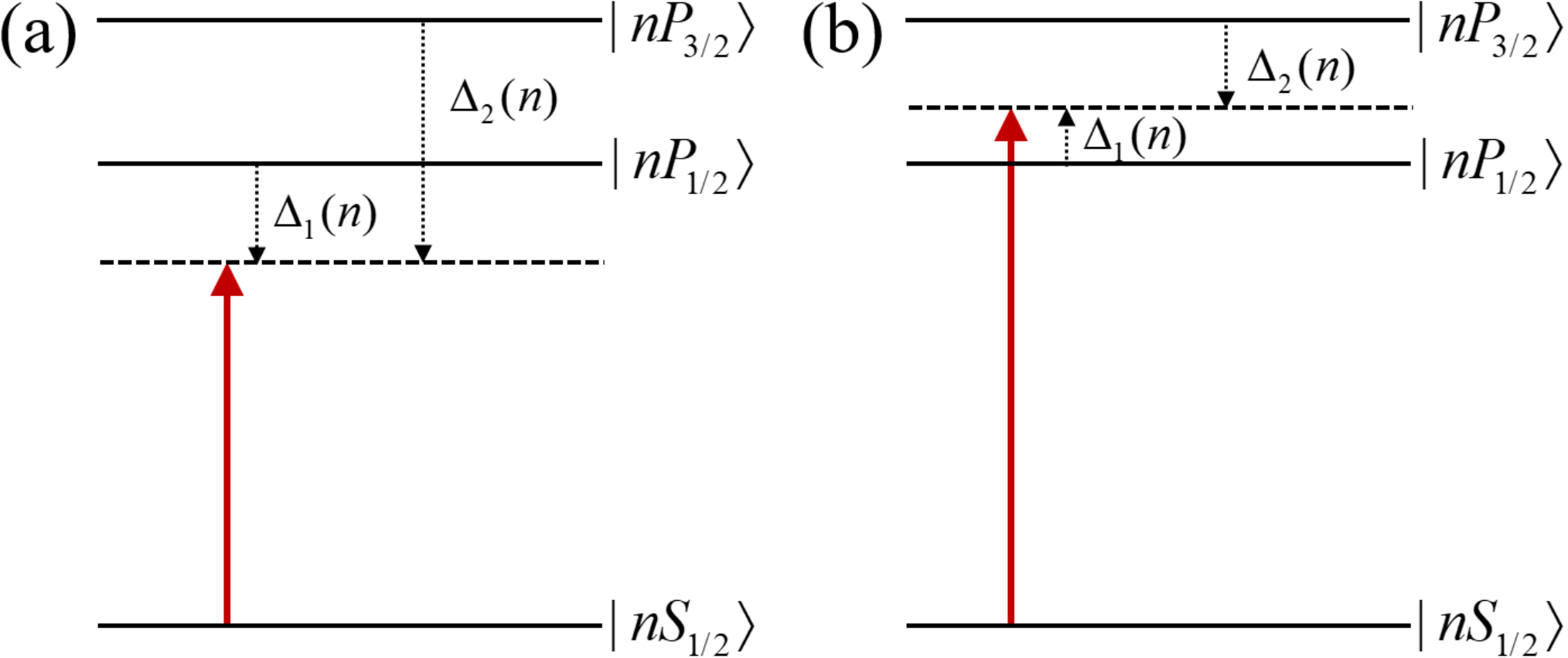}
	\caption   {\baselineskip 3.5ex 
		(a) A trap beam can be either red detuned from the $D_1$ transition
		or (b) tuned between the $D_1$ and $D_2$ transitions.}
\end{figure}

\begin{table}[t]
	\caption{ Parameters for analogous Zeeman effect. 
		$\lambda_0$ is the wavelength that satisfies Eq. (\ref{eq:analogous Zeeman}), 
		$\beta$ is the vector polarizability at $\lambda_0$.
		$\mathcal{B}_{\rm eff}$ and $R_\gamma$ are effective magnetic field and photon scattering rate, respectively, for a state with $g_Fm_F =1$
		when the laser intensity is $1.4 \times 10^9$ W/m$^2$
		with right-circular polarization. 
		\label{Table1}}
	\begin{tabular} {c c c c c} \hline \hline
		atom & \hspace{2 mm} $\lambda_0$ (nm) \hspace{2 mm} & $\beta$ (atomic units)  & \hspace{2 mm}  $\mathcal{B}_{\rm eff}$ (G) \hspace{2 mm}  & \hspace{2 mm} $R_\gamma$ (s$^{-1}$) \hspace{2 mm} \\ \hline
		Li & 670.8		& $5.4 \times 10^6$  &  $2.5 \times 10^4$ &   $2.0 \times 10^8$  \\ \hline
		Na& 589.4		& $1.2 \times 10^5$  &  $560$ &   $1.4 \times 10^5$  \\ \hline
		K &  768.8		& $4.8 \times 10^4$  &  $230$ &   $1.0 \times 10^4$  \\ \hline
		Rb & 789.8		& $1.2 \times 10^4$  &  $58$ &   $6.5 \times 10^2$  \\ \hline
		Cs & 879.8		& $6.0 \times 10^3$  &  $28$ &   $1.2 \times 10^2$  \\ \hline \hline
	\end{tabular} 
\end{table}

%\vspace{5 mm}

%\newpage
We experimentally confirmed the Zeeman-like ac Stark shift of Eq. (\ref{eq:ac Stark 4})
by performing stimulated Raman spectroscopy \cite{analogous Zeeman exp} on
a slow Rb beam leaked from a magneto-optical trap \cite{LVIS}. 
The intensity gradient of an elliptically polarized beam with a detuning of Eq. (\ref{eq:analogous Zeeman}) exerts an $m_F$-dependent force on an alkali-metal atom in the same way as a magnetic field gradient does.
The strength of the force is proportional to the degree of circularity $\eta$ in Eq. (\ref{eq:eta}).
We demonstrated that a spin-polarized Rb atom traversing an intensity gradient on one side of a Gaussian beam profile showed $\eta$-dependent deflection \cite{optical Stern-Gerlach}.
The experimental apparatus is shown in Fig. 2, and the results are shown in Fig. 3; 
for the right and left circularly polarized beam of $\eta =+1$ and $-1$, respectively, 
atoms are deflected in the opposite directions. 
We call it optical the Stern-Gerlach effect.
\begin{figure}[h] \centering
	\includegraphics[scale=0.56]{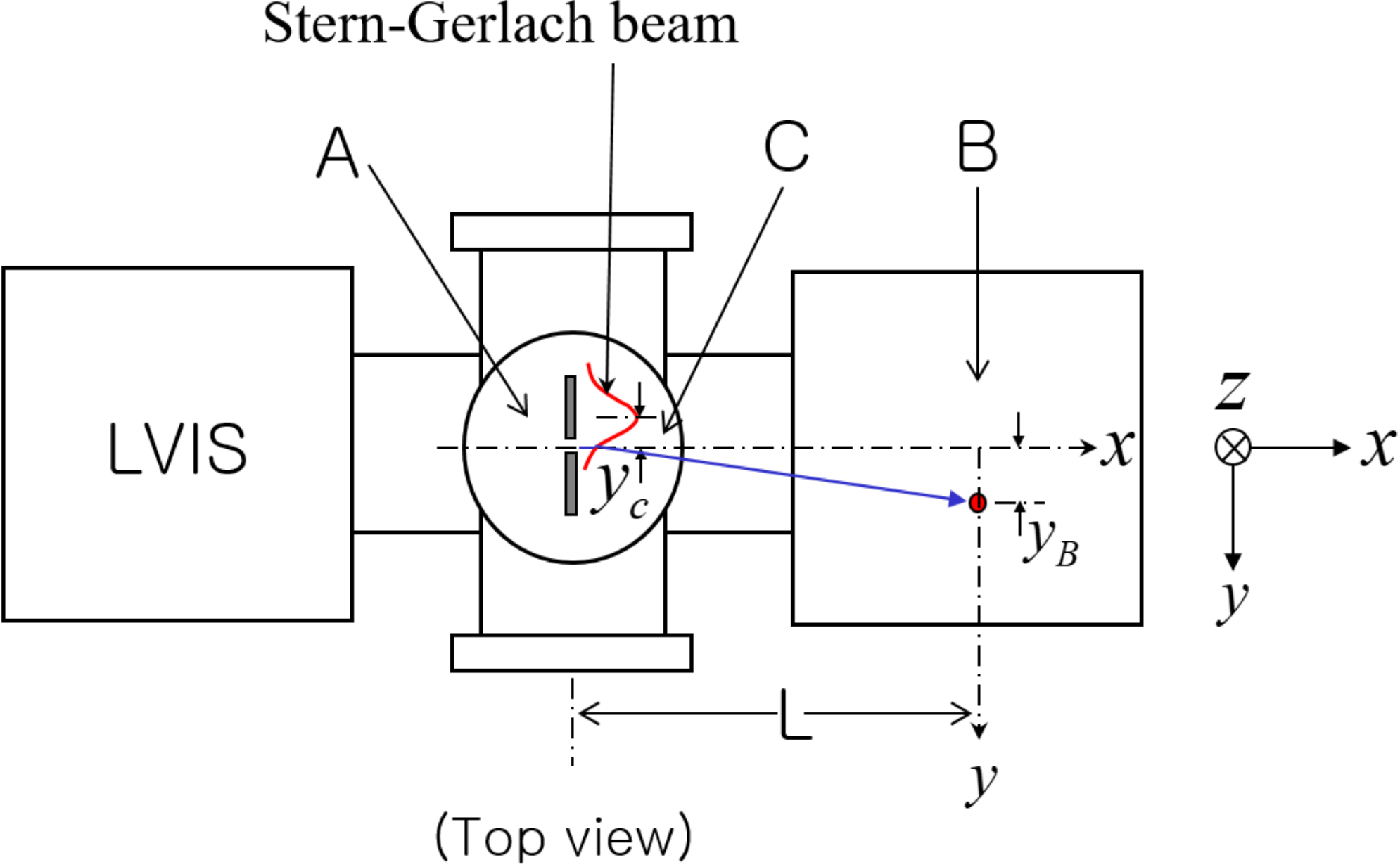}
	\caption   {\baselineskip 3.5ex 
		Apparatus for the optical Stern-Gerlach experiment. It consists of a
		low-velocity intense source of atoms, a spin-polarization region (A), 
		an interaction region (C), 		and a probe region (B).
		In the C region, a vertically propagating Gaussian
		beam is tightly focused.}
\end{figure}
\begin{figure}[h] \centering
	\includegraphics[scale=0.56]{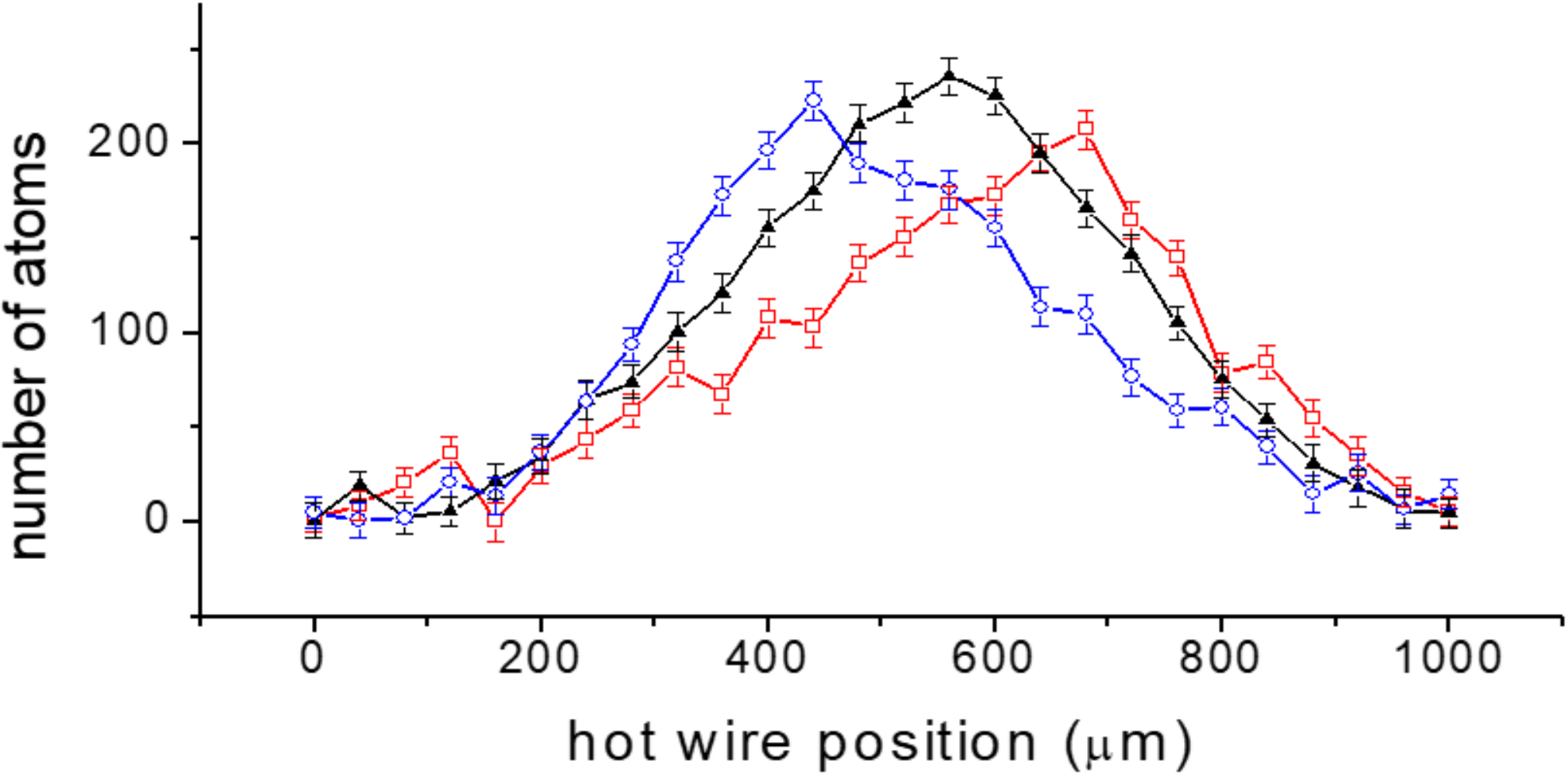}
	\caption   {\baselineskip 3.5ex 
		Hot-wire scan results. Triangles are when the Stern-Gerlach beam is blocked. Circles and squares are when the
		Stern-Gerlach beam is right and left circularly polarized,
		respectively.}
\end{figure}
While the term was used to describe the difference in radiation pressure experienced by 
two states of a fictitious spin associated with a two-level atom \cite{OSG old theory, OSG old exp},
our experiment is a more direct analogy to the original Stern-Gerlach experiment.
We also used the idea to create an optical trap that behaves like a magnetic trap  
and observed 98\%  spin polarization of the trapped atoms \cite{analogous magnetic trap}.
This idea has been applied to many interesting experiments.
As an example, in a quantum walk experiment in a one dimensional optical lattice, 
the optical Stern-Gerlach force by a circularly polarized lattice beam plays the role of a coin toss
on an atom in a superposition of opposite $g_Fm_F$ states \cite{quantum walk}.
Quantum gate operation by entangling two atoms in a three dimensional optical lattice
was realized by merging nearby sites by controlling lattice polarization \cite{Porto}.

%\newpage
\section{Magic polarization}
Motivation of the early research on laser cooling and trapping of atoms in the 1980s
was to increase interrogation time to improve precision in spectroscopy. 
However, an optically trapped atom suffers a state-dependent ac Stark shift, which
leads to inhomogeneous broadening of a transition, 
and most of the benefit of long interrogation time is lost. 
For an optical transition, such as the $D$ transitions of an alkali-metal atom,
the idea of magic wavelength can eliminate the broadening
when there is a proper three-level configuration \cite{magic wavelength}. 
For a ground hyperfine transition of an alkali-metal atom, 
the inhomogeneous broadening originates from different detuning for the $F$ and $F+1$ states
\begin{figure}[t] \centering
	\includegraphics[scale=0.55]{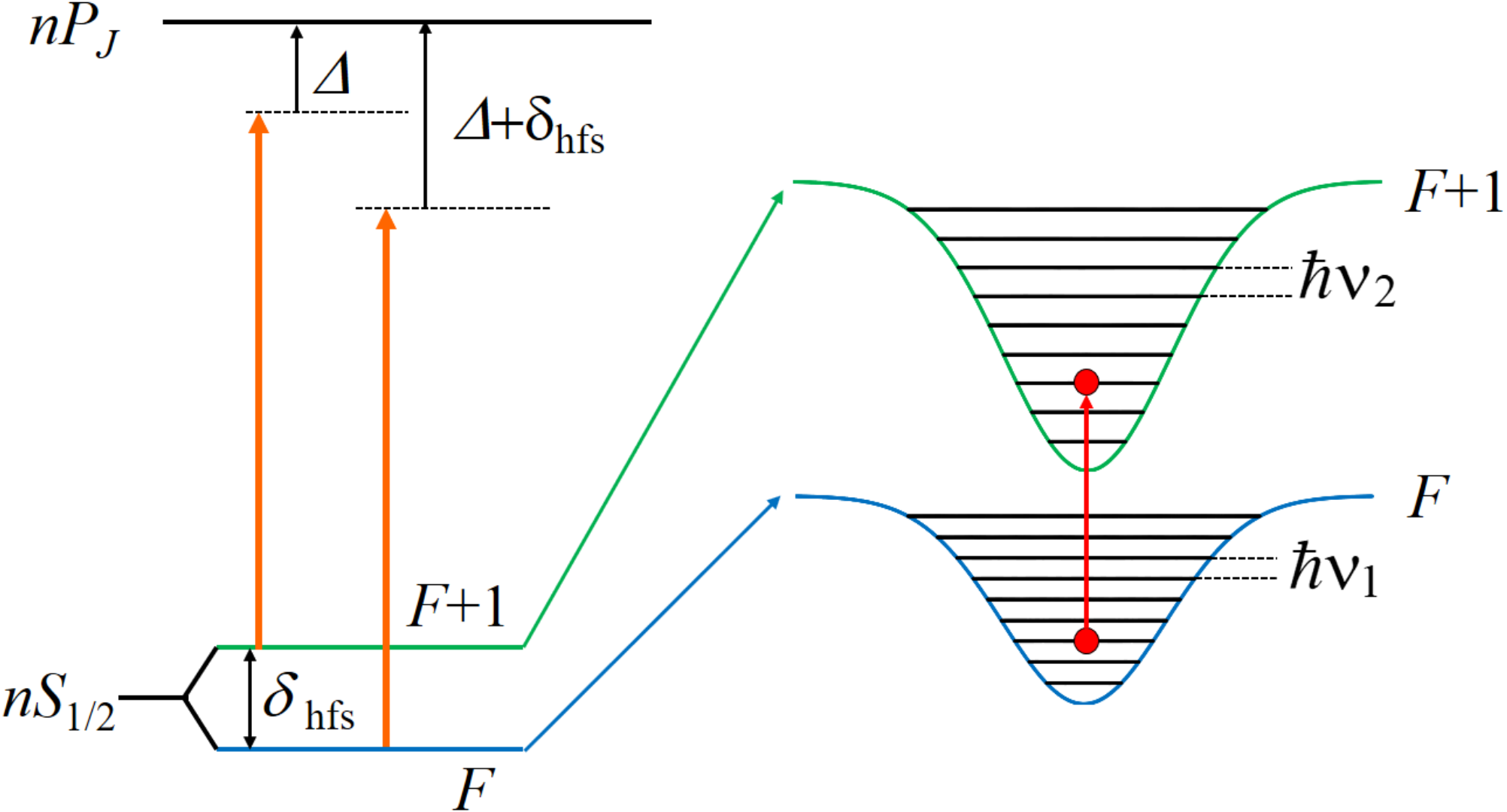}
	\caption   {\baselineskip 3.5ex 
		Inhomogeneous broadening of a ground hyperfine transition of an alkali-metal atom.}
\end{figure}
owing to the ground hyperfine splitting $\delta_{\rm hfs}$ as shown in Fig. 4.
Atoms in the $F$ and $F+1$ states have different ac Stark shifts, 
and consequently their well depths $U_1$ and $U_2$ and vibration frequencies $\nu_1$ and $\nu_2$, respectively, are different.
This implies that transition frequency between the vibrational states 
$|\chi_1(n)\rangle$ and $|\chi_2(n)\rangle$ of $U_1$ and $U_2$, respectively,
depends on the vibrational quantum number $n$,
and even a transition between $|\chi_1(n_1)\rangle$ and $|\chi_2(n_2)\rangle$ with $n_1 \neq n_2$ 
is possible.

%\newpage
Because the ground hyperfine transition plays an important role in precision measurement such as 
an atomic clock and search for symmetry violation, there were many discussions \cite{Meschede-inhomo}
and experimental efforts  to overcome this problem: for example,
use of an extra laser beam to compensate the differential shift \cite{two-beam1, two-beam2}, 
spin-echo or dynamic decoupling to undo Stark-induced dephasing \cite{spin echo},
and a blue-detuned optical bottle to minimize atom-light interaction \cite{blue bottle}. 
However, these approaches are designed only to reduce the broadening and dephasing.
In 2007, we proposed an idea that could eliminate the differential shift completely 
by having $U_1 = U_2$ for a transition of interest \cite{magic pol theory}.
The idea relies on the $\beta$-induced ac Stark shift in Eq. (\ref{eq:ac Stark 3}) again;
for a transition from the $|nS_{1/2}, F, m_F \rangle$ 
to the $|nS_{1/2}, F+1, m_F^\prime \rangle$ state,
the differential shift  vanishes 
when the degree of circularity satisfies
\begin{equation}
	\eta =\frac{\Delta \alpha}{\beta g_F (m_F +m_F^\prime)},
	\label{eq:magic pol condition}
\end{equation}
where $\Delta \alpha =\alpha_{F+1} -\alpha_F$ 
with $\alpha_{F+1}$ and $\alpha_F$ being the scalar polarizabilities of the $F+1$ and $F$ state, respectively, and 
$\beta$ is an average value of those for the $F+1$ and $F$ states.
We call it the magic polarization.
The same idea was published in 2008 by a theory group \cite{Flambaum}, 
who proposed to tune the angle $\theta$ between 
a quantization axis defined by a magnetic field and trap-beam propagation direction
to eliminate the differential ac Stark shift by noting that 
the effective degree of circularity is given by
\begin{equation}
	\eta_{\rm eff} = \eta \cos \theta.
	\label{eq:Flambaum}
\end{equation}
 
Although both proposals in Refs. \cite{magic pol theory} and \cite{Flambaum} considered 
a hyperfine transition of an optically trapped Cs atom,
it is not the best atom to apply the idea 
because  its large $\beta$ puts very stringent limit on polarization control. 
Optically trapped atoms at temperature $T$ have an average vibrational quantum number 
$\langle n \rangle = k_B T/\hbar \nu$ in a one-dimensional model,
where $k_B$ is the Boltzmann constant. 
When a polarization error results in $\delta \eta$ from the magic condition in Eq. (\ref{eq:magic pol condition}), frequency of a $\Delta n =0$ transition for $n = \langle n \rangle$ is shifted by 
$\delta \! f$   from that for $n=0$ with $\delta \eta$ and $\delta \! f$ being related by 
\begin{equation}
\delta \eta = \frac{\alpha}{\beta g_F(m_F+m_F^\prime)}\cdot\frac{h}{k_B T} \delta \! f.
\label{tolerance on eta}
\end{equation}
\begin{table}[b]
	\caption{Tolerance on polarization control for most abundant isotopes of alkali-metal atoms. $\lambda_{\rm trap}$ is the wavelength when a trap beam is 100 THz red detuned from the $D_1$ transition.
		$f_{\rm hf}$ is the ground hyperfine splitting.
		$\alpha$ and $\beta$  at $\lambda_{\rm trap}$ are given in atomic units. 
		$\eta_{\rm magic}$ satisfies the magic-polarization condition in Eq. (\ref{eq:magic pol condition}) 
		when $g_F(m_F+m_F^\prime)=1$, 
		and $\delta \eta_{\rm max}$ is tolerance on $\eta$ when $\delta \! f=1$ Hz and 
		atomic temperature is 100 $\mu$K.
		\label{Table2}}
	\begin{tabular} {c c c c c c c} \hline \hline
		atom & \hspace{2 mm} $\lambda_{\rm trap}$ (nm) \hspace{2 mm} & $f_{\rm hf}$ (GHz) 
		& $\alpha$ (a.u.)  & $\beta$ (a.u.) & \hspace{2 mm} $\eta_{\rm magic}$ \hspace{2 mm} &  \hspace{2 mm} $\delta \eta_{\rm max}$ \hspace{3 mm} \\ \hline
		$^7$Li & 864 & $0.803$& $407$  & 0.016 & $0.18$  & \hspace{2 mm} $1.2 \times 10^{-2}$ \\ \hline
		$^{23}$Na& 734 & $1.772$  & $453$ &   $0.66 $  & $1.1 \times 10^{-2}$ & \hspace{2 mm}$3.3 \times 10^{-4}$\\ \hline
		$^{39}$K & 1036 & $0.462$  &  $629$ & $3.3$ & $7.5 \times 10^{-4}$ & \hspace{2 mm}$9.2 \times 10^{-5}$\\ \hline
		$^{87}$Rb & 1082		& $6.835$  &  $652$ &   13  & $2.8\times 10^{-3}$ &\hspace{2 mm}$2.4 \times 10^{-5}$ \\ \hline
		$^{133}$Cs & 1275		& $9.193$  &  $714$ &   32  & $1.5\times 10^{-3}$ &\hspace{2 mm}$1.1 \times 10^{-5}$\\ \hline \hline
	\end{tabular} 
\end{table}
Any imperfection in polarization-control optics such as a linear polarizer or a retardation plate and 
residual birefringence of a lens or a window can introduce $\delta \eta$.
We note that for a given $\delta \! f$, tolerance $\delta \eta_{\rm max}$ on polarization control is inversely proportional to $\beta$, 
and it becomes small for heavy atoms. 
In Table II, $\delta \eta_{\rm max}$ for $\delta \! f=1$ Hz is given 
when a trap beam is 100 THz red detuned from the $D_1$ transition, $T=100$ $\mu$K, and
$g_F(m_F+m_F^\prime)=1$. 
For experimental demonstration of magic polarization, we used the
$|2S_{1/2}, F=1, m_F=1 \rangle$ to $|2S_{1/2}, F=2, m_F=2 \rangle$ transition of 
$^7$Li atoms in a 1064-nm optical trap \cite{magic pol experiment}.
Fig. 5(a) and (b) show full width at half maximum (FWHM) of Rabi lineshape and coherence time $\tau_c$ from Ramsey spectroscopy, respectively, versus $\eta$. 
\begin{figure}[t] \centering
	\includegraphics[scale=0.8]{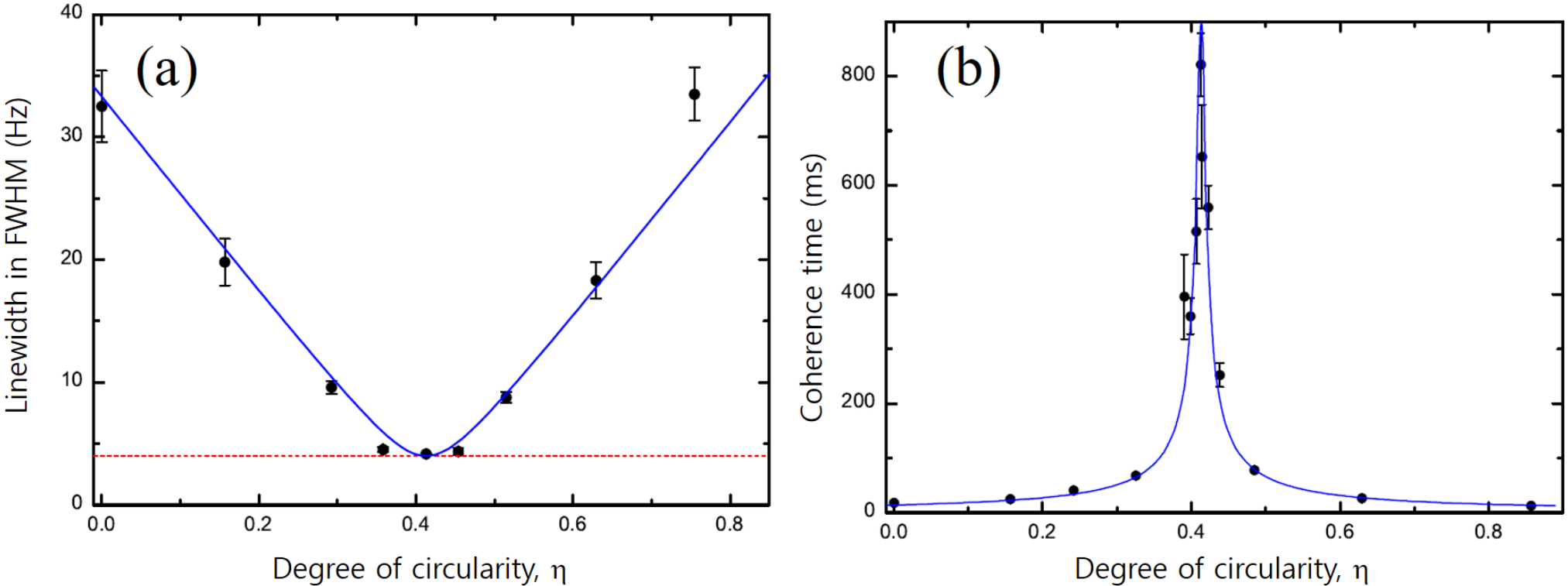}
	\caption   {\baselineskip 3.5ex 
		(a) Full width at half maximum when rf field is applied for 200 ms 
		and (b) coherence time of the 
		$|2S_{1/2}, F=1, m_F=1 \rangle$ to $|2S_{1/2}, F=2, m_F=2 \rangle$ transition of 
		$^7$Li atoms in a 1064-nm optical trap versus degree of circularity $\eta$. }
\end{figure}
We observed minimum linewidth and maximum $\tau_c$ at $\eta = 0.413$ 
while $\eta_{\rm magic}$ calculated from the published data \cite{Safronova} is 0.39. 
When we applied 803-MHz field for 2 s, 
FWHM was as small as $0.59 \pm 0.02$ Hz, while the Fourier-limited value was 0.4 Hz.
The maximum $\tau_c$ in Fig. 5(b) was $820 \pm 60$ ms, largely limited by magnetic field noise.

The narrow linewidth allowed us to address individual Li atoms in a 1D optical lattice 
with a site-specific resolution, and 
the long coherence time allowed us to coherently manipulate individual atoms independently of neighboring ones \cite{PRL 2019}.
Figure 6 shows Rabi signal from single atoms at three adjacent sites versus rf frequency
while a magnetic field gradient along the lattice introduces 
differential Zeeman shift $\Delta f_{\rm Zeeman}$ of 180 Hz between neighboring sites. 
Resolving power in the frequency domain, 
defined as $\Delta f_{\rm Zeeman}$ divided by FWHM of each Rabi lineshape, is  2.3.
It is three times that in the spatial domain  of 0.8 
obtained by a state-of-the-art quantum-gas microscope on Li atoms in an optical lattice \cite{Choi KAIST}.
Here, the resolving power in the spatial domain is defined as half the lattice wavelength divided by FWHM of 
a point spread function of an imaging optics.
\begin{figure}[t] \centering
	\includegraphics[scale=0.45]{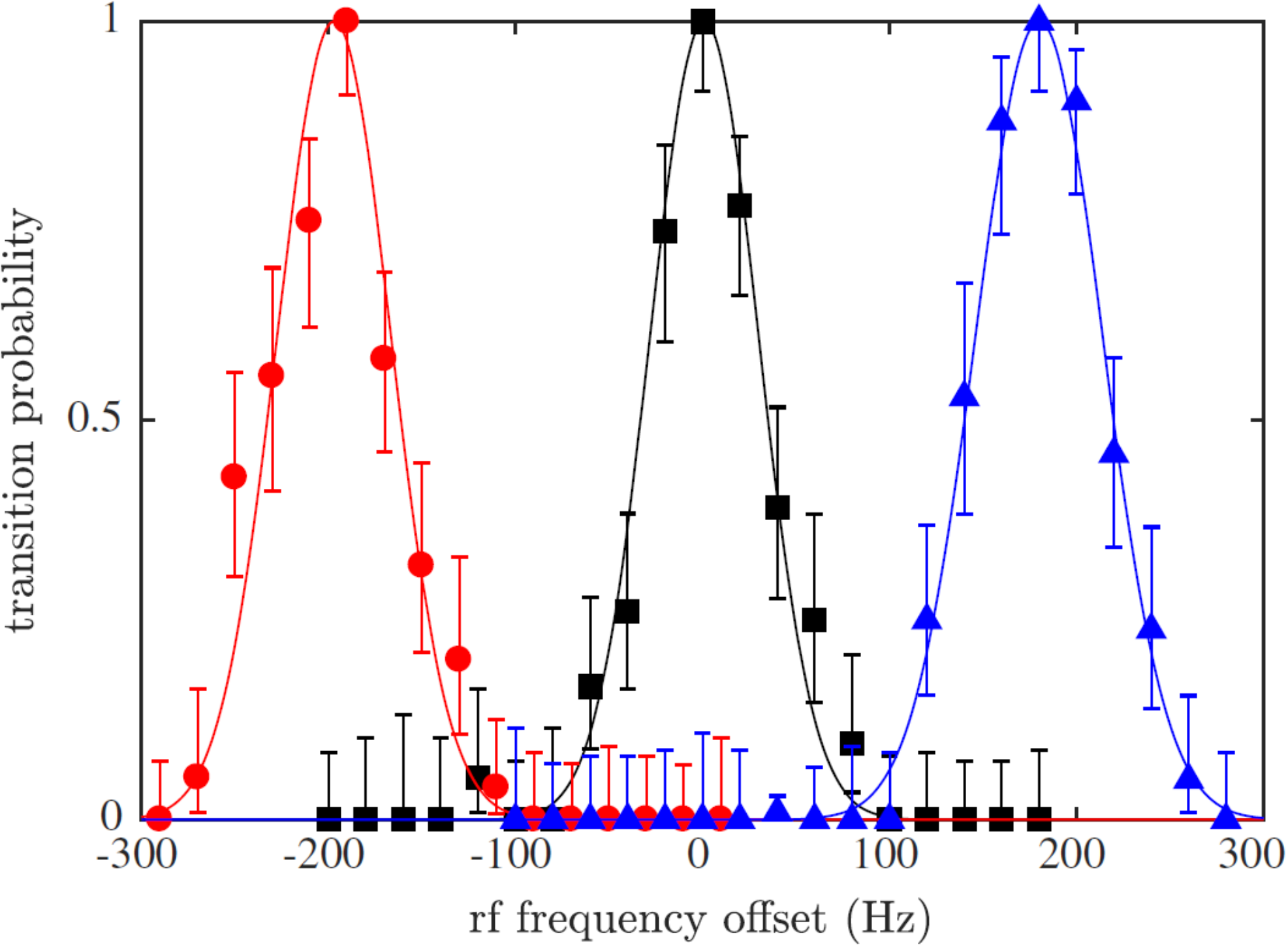}
	\caption   {\baselineskip 3.5ex 
		Rabi line shapes for single atoms residing at the $q_{-1}$
		(red circle), $q_0$ (black square), and $q_{+1}$ sites (blue triangle).
		Magnetic field gradient is 1.6 G/cm. Solid curves are theory
		predictions for a $\pi$ pulse driven by a 20-ms Blackman window.}
\end{figure}

For a clock transition between $m_F=0$ states, where the first-order Zeeman shift vanishes and 
the idea of magic polarization is not applicable, 
Derevianko proposed to use the second-order shift to eliminate the inhomogeneous broadening
\cite{Derevianko}.
When $\vec{B} = B_0\hat{z}$ is applied, the second-order Zeeman shift of a clock transition at $f_0$ is
\begin{equation}
	\Delta f_Z^{(2)}\approx 2 f_0 \left( \frac{\mu_B B_0}{hf_0}\right)^2.
	\label{eq:second-order Zeeman}
\end{equation}
In addition, inside a right circularly polarized trap beam, 
the total shift $\Delta f_0$ of the clock transition is given by
\begin{equation}
	h \Delta f_0 = \Delta \alpha |\vec{\mathcal{E}}|^2
	+\frac{2{\mu_B}^2}{hf_0} \left( B_0-\frac{\beta}{\mu_B}|\vec{\mathcal{E}}|^2\right)^2,
\end{equation}
where $\Delta \alpha =\alpha_{F+1} -\alpha_F$ and 
$B_0$ in Eq. (\ref{eq:second-order Zeeman}) is replaced by $B_0+\mathcal{B}_{\rm eff}$.
Inhomogeneous broadening owing to $\Delta \alpha$ is eliminated when 
\begin{equation}
	B_0 = \frac{\Delta \alpha}{\beta} \cdot \frac{h f_0}{4\mu_B },
\end{equation}
which may be called a magic B field. 
Effect from $|\mathcal{B}_{\rm eff}|^2$ is negligible for an alkali-metal atom
 owing to the smallness of $|\mu_B \mathcal{B}_{\rm eff}/hf_0|$ in a typical optical trap.

%\newpage
\section{Magic well depth}
One of the most effective ways to cool trapped atoms is Raman sideband cooling (RSC) \cite{RSC Wineland}.
We consider RSC applied to alkali-metal atoms trapped in a 1D optical lattice formed by a standing wave 
of a Gaussian beam with a minimum spot size $w_0$ and wavelength $\lambda_{\rm OL}$.
Its potential well is 
\begin{equation}
	U(\vec{r} \,) = U_0 e^{-2(x^2+y^2)/w_0^2} \cos^2 k z,
	\label{eq:U(r)}
\end{equation}
where $U_0=U_{\rm{AC}} (F,m_F)$ in Eq. (\ref{eq:ac Stark 3}) and $k=2\pi/\lambda_{\rm OL}$.
We assume a large Rayleigh range and therefore a constant spot size.
In a harmonic approximation, the state of a lattice-bound atom is described by
$|\psi_j (\mathbf{n})\rangle = |\phi_j\rangle \otimes |\chi(\mathbf{n}) \rangle$,
where $|\phi_j \rangle$ is one of the ground hyperfine states and
$|\chi(\mathbf{n}) \rangle$ with vibrational quantum numbers $\mathbf{n}=(n_x, n_y, n_z)$ 
represents a 3D harmonic oscillator.
Longitudinal and transverse vibration frequencies are 
$\nu_z = k \sqrt{2|U_0|/m}$ and $\nu_\bot = (2/w_0) \sqrt{|U_0|/m}$, respectively,
with $m$ being the atomic mass. 
For RSC along the $z$ axis, we use a Raman transition 
$|\psi_1 (\mathbf{n})\rangle \rightarrow |\psi_2 (\mathbf{n}^\prime)\rangle$ 
with $|\phi_1 \rangle = |nS_{1/2}, F, m_F \rangle$,  
 $|\phi_2 \rangle = |nS_{1/2}, F+1, m_F^\prime \rangle$, and 
$n_x = n_x^\prime$, $n_y = n_y^\prime$.
The transition frequency $\omega_R (\mathbf{n}, \mathbf{n^\prime})$ is 
$\omega_{12} + \Delta n_z \nu_z$, where
$\omega_{12}$ is for the $|\phi_1 \rangle \rightarrow |\phi_2 \rangle$ transition of a free atom 
 and  $\Delta n_z = n_z^\prime - n_z$.

In practice, the Raman transition suffers an inhomogeneous broadening 
from anharmonicity of the potential well as well as from
the usual differential ac Stark shift between $|\phi_1 \rangle$ and $|\phi_2 \rangle$.
For anharmonicity, we consider the quartic terms from $U(\vec{r} \,)$:
\begin{equation}
	W_q =U_0 \left( \frac{1}{3}k^4z^4   + 2\,\frac{k^2\rho^2 z^2}{w_0^2}  + 2\frac{\rho^4}{w_0^4}  \right),
	\label{eq:quartic}
\end{equation}
where $\rho^2 =x^2+y^2$.
While the carrier frequency is not shifted by $W_q$ in the first-order perturbation, 
the sidebands for $\Delta n_z =\pm1$ suffer a shift by $\Delta \omega_q (\mathbf{n}, \mathbf{n^\prime})$:
\begin{equation}
	\Delta \omega_q (\mathbf{n}, \mathbf{n^\prime}) = 
	-\Delta n_z \left\{  \Delta \omega_q^{(1)} \left( n_z +\frac{1}{2} \right)+     \Delta \omega_q^{(2)}(n_\bot+1)         \right\},
	\label{eq:shiftq}
\end{equation}
where $n_\bot = n_x +n_y$,
$ \Delta \omega_q^{(1)} = \hbar k^2 /2m,$  and $\Delta \omega_q^{(2)} = k \hbar / \sqrt{2} mw_0$.
We neglect an offset term which is independent of $\mathbf{n}$ and $\mathbf{n}^\prime$.
In addition, from the dependence of $U_0$ on $F$ and $m_F$ in Eq. (\ref{eq:ac Stark 3}), 
$\omega_R (\mathbf{n}, \mathbf{n^\prime})$ is further shifted by
$\Delta \omega_\beta (\mathbf{n}, \mathbf{n^\prime})$:
\begin{equation}
	\Delta \omega_\beta (\mathbf{n}, \mathbf{n^\prime}) =
\Delta \omega_\beta^{(1)} \left( n_z +\frac{1}{2} \right) 
	+  \Delta \omega_\beta^{(2)}(n_\bot+1),
	\label{eq:shiftb}
\end{equation}
where $\Delta \omega_\beta^{(1)} = \{\eta\beta g_F(m_F+m_F^\prime)/2\alpha\} \omega_z$ and 
$\Delta \omega_\beta^{(2)} = \{\eta\beta g_F(m_F+m_F^\prime)/2\alpha\} \omega_\bot$.
Here we assume $|\Delta \alpha| \ll |\eta\beta g_F(m_F+m_F^\prime)|$, 
which is valid for a heavy alkali-metal atom in a circularly polarized trap. 
From Eqs. (\ref{eq:shiftq}) and (\ref{eq:shiftb}), we note that the total shift 
$\Delta \omega_q (\mathbf{n}, \mathbf{n^\prime}) +\Delta \omega_\beta (\mathbf{n}, \mathbf{n^\prime})$
vanishes if 
$\Delta \omega_\beta^{(1)} =\Delta n_z \Delta \omega_q^{(1)}$ and
$\Delta \omega_\beta^{(2)} =\Delta n_z \Delta \omega_q^{(2)}$.
Because the ratio
$\Delta \omega_q^{(1)} /\Delta \omega_q^{(2)}$ is the same as 
$\Delta \omega_\beta^{(1)} /\Delta \omega_\beta^{(2)}$, 
the two conditions are simultaneously satisfied when
\begin{equation}
	|U_0| =\left[\frac{\alpha}{\beta g_F(m_F+m_F^\prime)} \right]^2 \! \!\! \cdot \frac{\hbar^2 k^2}{2m}
	\label{eq:magic well depth}
\end{equation}
and $\eta =\Delta n_z$. 
We may call $|U_0|$ a magic well depth. 
By using either a right or a left circularly polarized beam, the total shift can be eliminated for
either blue or red sideband, respectively. 
If a trap beam is elliptically polarized with $|\eta| <1$, 
the magic $|U_0|$ increases by $1/\eta^2$, giving flexibility in designing an experiment. 

\begin{figure}[b] \centering
	\includegraphics[scale=0.43]{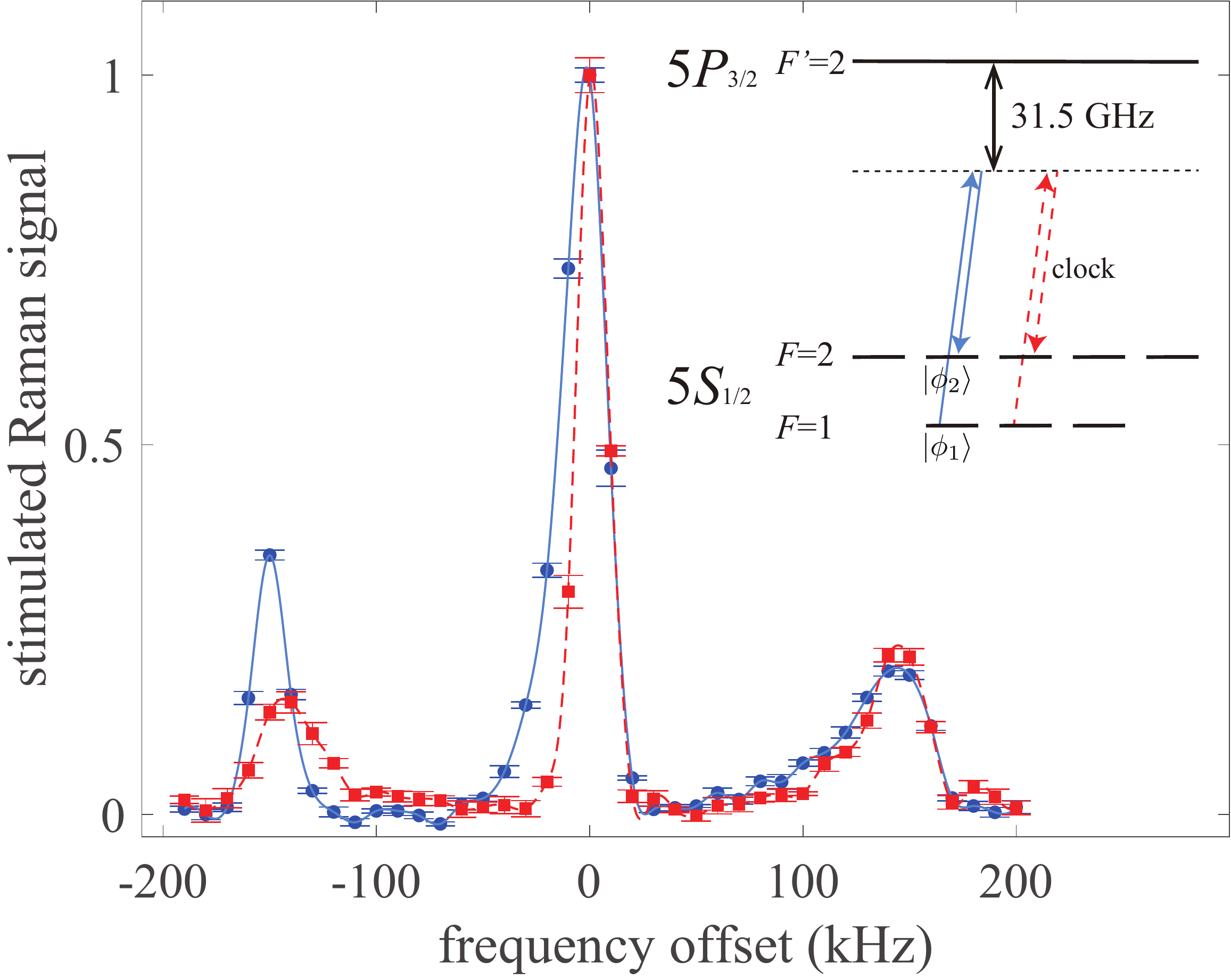}
	\caption   {\baselineskip 3.5ex 
		Stimulated Raman spectrum at the magic well depth. 
		Red sideband of the $ |5S_{1/2}, F=1, m_{_F}=-1 \rangle$ $\rightarrow$ $ |5S_{1/2}, F=2, m_{_F}=-1 \rangle$ transition (solid line with blue circles) is narrowed by cancellation of broadening from 
		anharmonicity and vector polarizability. 
		Spectrum of the clock transition (dashed line with red squares), which suffers broadening from only the anharmonicity, is shown for comparison. 
		Each spectrum is normalized to its carrier signal.  }
\end{figure}

We have experimentally demonstrated the idea of magic well depth
by eliminating inhomogeneous broadening of a Raman sideband transition 
using $^{87}$Rb atoms trapped in a 1D optical lattice \cite{magic well depth}.
Figure 7 shows the lineshape of simulated Raman transitions:
blue circles are for the transition of interest,
 $|5S_{1/2}, F=1, m_F=-1 \rangle$ $\rightarrow$ $|5S_{1/2}, F=2, m_F^\prime=-1 \rangle$, 
 and red squares are for the clock transition, 
whose broadening is only from the anharmonicity.  
We use a left circularly polarized lattice beam at $\lambda_{\rm OL} = 980$ nm 
and the magic well depth of $|U_0|/k_B= 140$ $\mu$K, 
determined by $\lambda_{\rm OL}$ and the transition of interest. 
Focusing on the red sideband of both transitions, we observe the line-narrowing effect of magic well depth.
The carrier of the clock transition is the narrowest because it is free from both 
$	\Delta \omega_q (\mathbf{n}, \mathbf{n^\prime})$
and 	$\Delta \omega_\beta (\mathbf{n}, \mathbf{n^\prime})$.

\section{motion-selective coherent population trapping}
RSC puts trapped atoms to the motional ground state with high probability only when 
the Lamb-Dicke parameter $\eta_{\rm  LD} =\sqrt{\mathcal{E}_R /\hbar\nu}$ is much smaller than 1.
Here, $\mathcal{E}_R$ is the recoil energy accompanying the emission of a photon and
$\nu$ is the vibrational frequency of a trapped  atom.
While the condition is well satisfied along tightly confined directions in an optical lattice 
or an optical tweezer, it is difficult to satisfy the condition in a typical optical trap. 
To improve the effectiveness of RSC outside the Lamb-Dicke regime, we proposed 
to incorporate the phenomenon of coherent population trapping (CPT)	to RSC \cite{MSCPT theory}. 
Figure 8 shows the inverted $\mathsf{Y}$ configuration for the cooling scheme. 
A usual RSC (shown  in black) consists of a red-detuned Raman transition $|\phi_1 \rangle \rightarrow$ $ |\phi_3 \rangle$,
which we call $p$ transition,  and 
the $D$ transition $|\phi_3 \rangle \rightarrow |\phi_4 \rangle$ for optical pumping. 
To incorporate CPT, we add $q$ Raman transition $|\phi_2 \rangle \rightarrow |\phi_3 \rangle$
 (shown in red) to form a $\Lambda$ configuration. 
To make the CPT resonance condition dependent on the motional quantum number $n$,
we use a circularly polarized trap beam so that $\nu_1 \neq \nu_2$ owing to the vector polarizability.
Here, $\nu_1$ and $\nu_2$ are the vibration frequencies of the motional states,
$|\chi_1\rangle$ and $|\chi_2 \rangle$, 
for the potential wells of $|\phi_1 \rangle$ and $|\phi_2 \rangle$, respectively.
Their difference is
\begin{equation}
	\Delta\nu_{12} =\frac{\beta}{4\alpha}\,\nu_0,
	\label{eq: Delta nu 12}
\end{equation}
where $\nu_0$ is the vibration frequency in a linearly polarized trap.
When the $p$ and $q$ pairs of Raman beams are tuned to 
the $\Lambda$ resonance between the $|\phi_1, \chi_1(0) \rangle$ and $|\phi_2, \chi_2(0) \rangle$ states,
the motional ground states form a CPT dark state:
\begin{equation}
	|\Psi (\rm CPT)\rangle = \frac{ |\phi_1, \chi_1(0) \rangle -|\phi_2, \chi_2(0) \rangle}{\sqrt{2}},
	\label{eq: CPT dark state}
\end{equation}
where atoms are accumulated.
\begin{figure}[t] \centering
	\includegraphics[scale=0.4]{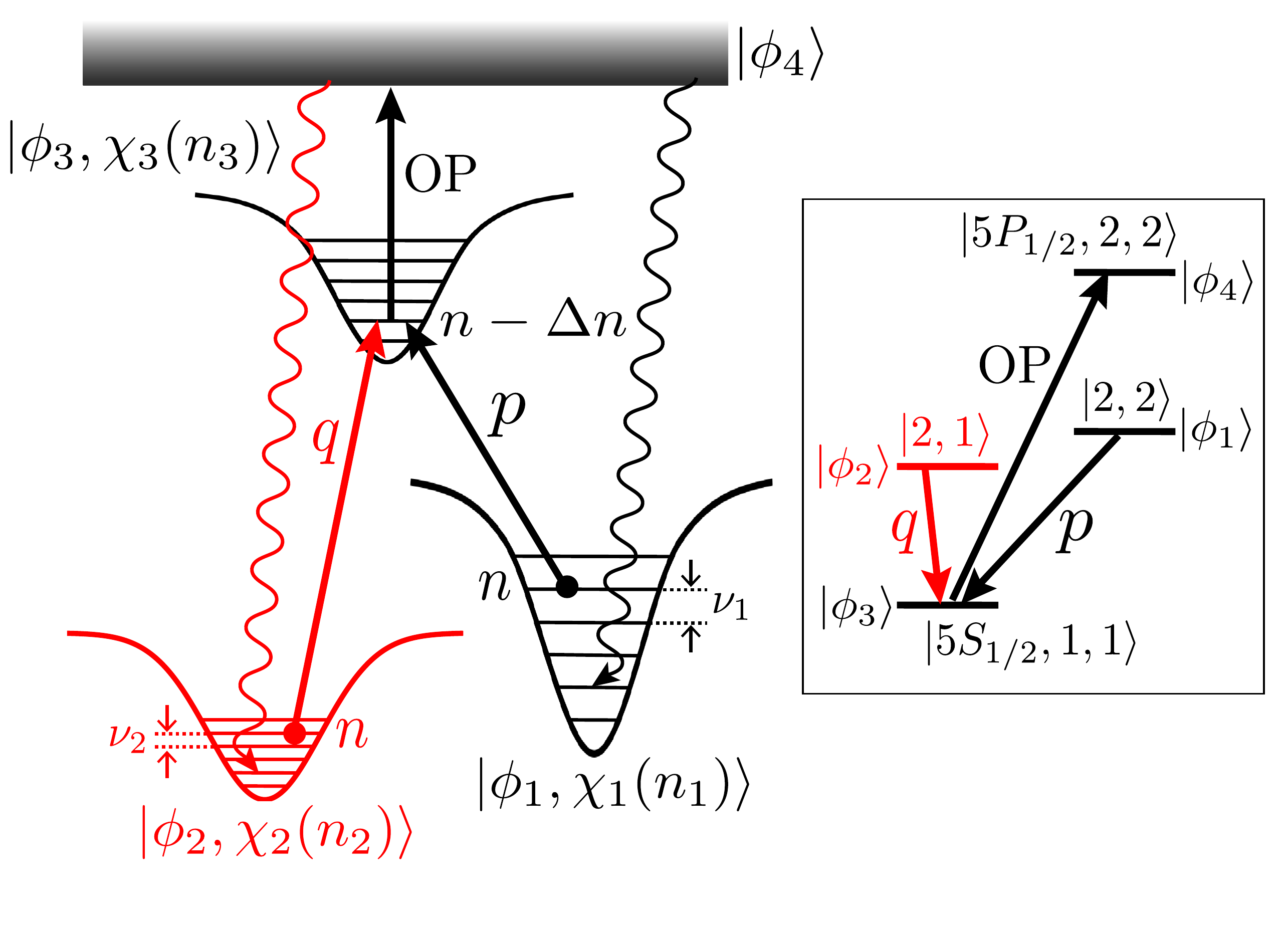}
	\caption   {\baselineskip 3.5ex 
		Level schemes for RSC and inverted $\mathsf{Y}$ configuration of MSCPT experiment. 
		RSC scheme is shown in black and $q$ transition of MSCPT is shown in red.
		Specific levels and transitions in $^{87}$Rb are assigned in the inset.}
\end{figure}
A pair of $|\phi_1, \chi_1(n) \rangle$ and $|\phi_2, \chi_2(n) \rangle$ states
are detuned from the CPT resonance by $n\Delta\nu_{12}$,
and a pair with the larger $n$ is brighter.
We call this method motion-selective coherent population trapping (MSCPT), and it is a
trapped-atom analogue to the velocity-selective CPT developed for
subrecoil cooling of free He atoms \cite{VSCPT 1988}.
Heavy alkali-metal atoms with large $\beta$ are favored for MSCPT.
For a $^{87}$Rb atom, if the trap wavelength is 850 nm and $\eta_{LD}=1$,
$\Delta\nu_{12}/ 2\pi = 75$ Hz, 
while FWHM of the CPT resonance in our radio frequency (rf) experiment was 150 Hz \cite{M1 CPT}.
In the inset of Fig. 8, specific levels in $^{87}$Rb are assigned: 
$|\phi_1 \rangle = |5S_{1/2}, F= 2, m_F = 2 \rangle$, 
$|\phi_2 \rangle =|2, 1 \rangle$,
$ |\phi_3 \rangle= |1, 1 \rangle$,
and $|\phi_4 \rangle=|5P_{1/2}, 2, 2 \rangle$.
We note that the inverted $\mathsf{Y}$ configuration is closed owing to the angular momentum 
selection rule and, as a consequence,  recoil heating during optical pumping is minimized.

Recently, we demonstrated the MSCPT cooling scheme in an experiment 
using $^{87}$Rb in a 1D optical lattice \cite{MSCPT exp}.
The lattice parameters were $w_0= 50$ $\mu$m, $\lambda_{\rm OL} = 980$ nm, and 
$U_0/k_B =125$ $\mu$K, resulting in $\eta_{\rm LD} = 2.3$ and 
$\Delta\nu_{12}/2\pi = 5$ Hz along the transverse direction.
Although the parameters were not optimal for MSCPT 
because we used an apparatus originally built for a different experiment,
we observed significant improvement in cooling efficiency compared with RSC
when the detuning $\delta_{\rm CPT}(0)$ of the $p$ and $q$ Raman beams 
from the CPT resonance of the $n=0$ states was close to 0.
In Fig. 9, temperature of the atoms in a steady state versus $\delta_{\rm CPT}(0)$ is shown for
MSCPT (blue circles) and RSC (black triangles).
To demonstrate the full benefit of the MSCPT scheme, 
we are upgrading the apparatus so that $\eta_{\rm LD} \simeq 1$ and $\Delta \nu_{12} \ge 75$ Hz
by using $w_0= 16$ $\mu$m, $\lambda_{\rm OL} = 850$ nm.
Finally, we note that a diatomic polar molecule in an optical trap \cite{polar molecule}
provides an excellent opportunity to apply MSCPT 
because its Stark shift depends strongly on the rotational quantum number.
Considering a MgF molecule in a 532-nm optical trap as an example, the fractional difference
between vibration frequencies of a pair of states in the same ro-vibrational level can be
as large as 12\%.
This may be compared with $\Delta \nu_{12}/\nu_0$ of less than 2\% for $^{87}$Rb in the new design.

\vspace{3 mm}
\begin{figure}[h] \centering
	\includegraphics[scale=0.55]{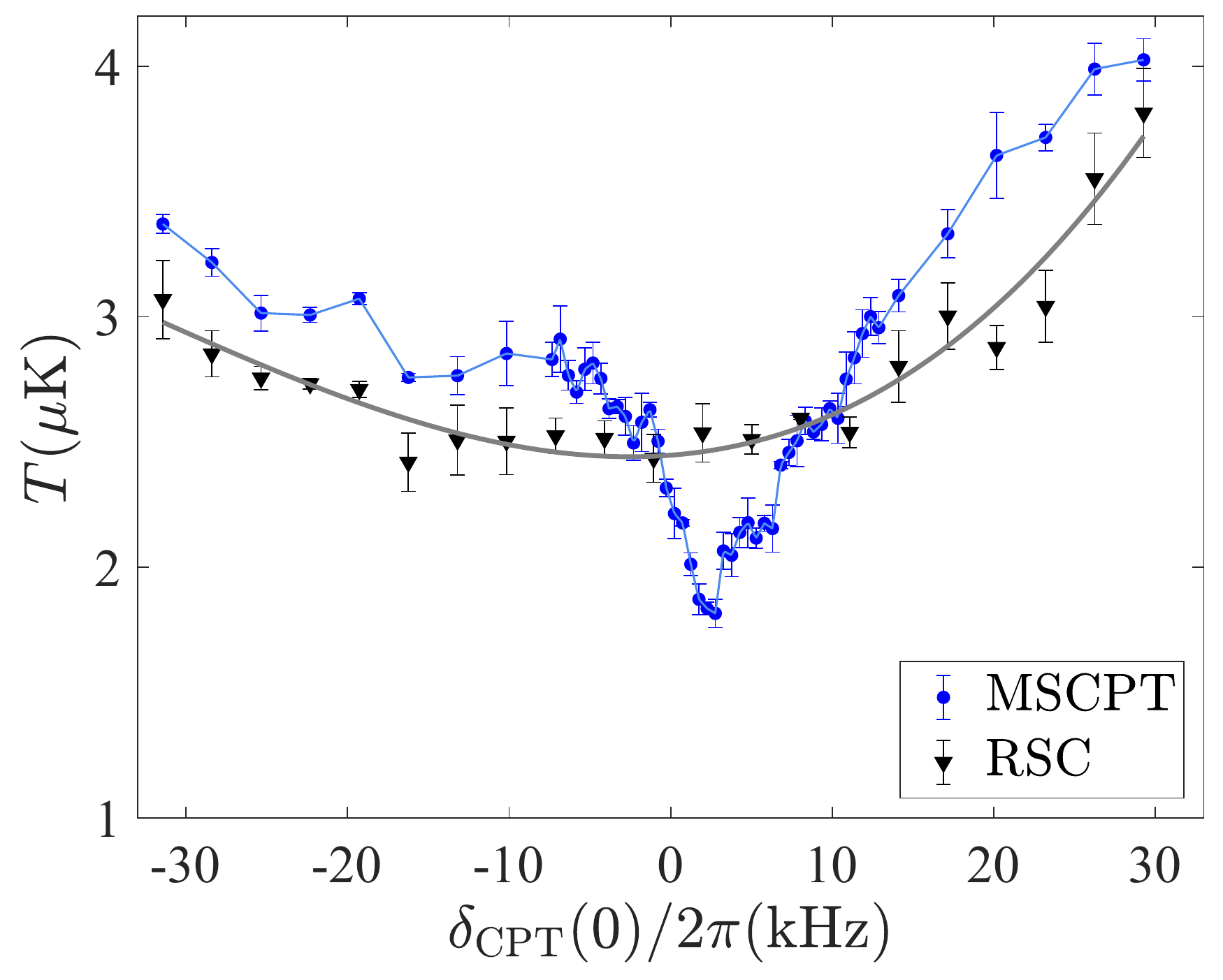}
	\caption   {\baselineskip 3.5ex 
	Temperature versus detuning $\delta_{\rm CPT}(0)$ from the CPT resonance for the motional ground state.
	Blue circles represent results from MSCPT cooling scheme, and 
	black triangles represent results when the $p$ and $q$ Raman  beams are alternately 
	turned on at 1 kHz.}
\end{figure}

\section{conclusion}
Over the last two and a half decades, our laboratory at Korea University 
has proposed several ideas using vector polarizability to manipulate alkali-metal atoms and 
carried out experiments to demonstrate those ideas. 
In this paper, we reviewed the ideas and experimental results.
Currently, we are working on an experiment to probe the cooling limit of a scheme 
based on motion-selective coherent population trapping 
and study strategies to apply the scheme to diatomic polar molecules in an optical trap.

\newpage
\section*{ACKNOWLEDGMENTS}
This work was supported by the National Research Foundation of Korea (Grant No. NRF-2022R1F1A1075131).

\newpage
\appendix
\section{Numerical calculation of scalar and vector polarizabilities}
In the Excel file of the Supplementary Material, 
to obtain numerical values of $\alpha$ and $\beta$, we use the definitions in Eq. (\ref{eq:alpha}) and 
Eq. (\ref{eq:beta}), respectively, including the off-resonant terms.
Those terms are not negligible when a trap beam is far detuned.
When the principal quantum number of the ground state of an alkali-metal atom is $n$, 
we include couplings to only the $nP_{J^\prime}$ and $(n+1)P_{J^\prime}$ states 
with $J^\prime=1/2,3/2$.
To obtain numerical values of $a$ and $b$,
we use Eq. (\ref{eq:a}) and Eq.(\ref{eq:b}), respectively,
and restrict the sums to $n^\prime =n$. 
Because we consider a far detuned trap beam, either ground or excited state hyperfine splitting is 
not important, and the effect of an isotope shift is negligible too.

There are two definitions for a reduced matrix element commonly used in the literature. 
We follow that of Racah \cite{Racah}, 
according to which the Wigner-Eckart theorem is written as \cite{Edmonds}
\begin{equation}
  \langle n^\prime , J^\prime, m_J^\prime | d_q | n, J, m_J \rangle =
  (-1)^{J^\prime-m^\prime}
  \left(   \begin{array}{ccc} J^\prime & 1 & J \\ -m_J^\prime & q & m_J  \end{array} \right)
  \langle n^\prime J^\prime || d || n J \rangle,
\end{equation}
and the reduced matrix element for an electric dipole moment 
between the $|n^\prime P_{J^\prime} \rangle$ and $|nS_{1/2} \rangle$ states of an alkali-metal atom
is related to the lifetime $\tau(n^\prime P_{J^\prime})$ by
\begin{equation}
	|\langle n^\prime P_{J^\prime} || d || n S_{1/2} \rangle|^2
	= \frac{1}{\tau(n^\prime P_{J^\prime})} 
	\frac{3\pi\epsilon_0\hbar c^3}{\omega_0^3}(2J^\prime+1),
\end{equation}
where $\omega_0$ is the $|nS_{1/2} \rangle$ $\rightarrow$ $|n^\prime P_{J^\prime} \rangle$
 transition frequency.
For spectroscopic data, we use NIST Atomic Spectra Database Lines Data \cite{NIST},
where the Einstein coefficient $A(n^\prime P_{J^\prime}) = 1/\tau(n^\prime P_{J^\prime})$ 
of the upper state $|n^\prime P_{J^\prime} \rangle$
and the wavelength $\lambda_0 = 2\pi c /\omega_0$ are tabulated. 

In the Excel file, for a given wavelength $\lambda$ of a trap beam, 
$\alpha, \beta$ and $a,b$ are calculated in atomic units.
For a given power $P$, minimum spot size $w_0$  ($e^{-2}$ intensity radius), 
and degree of circularity $\eta$ of the trap beam, 
the well depth $U_{\rm AC}(F,m_F)$ for the ground hyperfine state $|nS_{1/2}, F, m_F \rangle$
according to Eq. (\ref{eq:ac Stark 3}) is given (in mK) by
\begin{equation}
	U_{\rm{AC}} (F,m_F)= (\alpha-\eta\beta g_F m_F)(c\mu_0 I_0/2).
	\label{eq:ac Stark A}
\end{equation}
The photon scattering rate $R_\gamma (F,m_F)$ according to Eq. (\ref{eq:R_gamma 2}) is given (in s$^{-1}$) by
\begin{equation}
	R_\gamma (F,m_F)= (a-\eta b g_F m_F)(c\mu_0 I_0/2)\Gamma,
	\label{eq:R_gamma A1}
\end{equation}
where we use $\Gamma = A(nP_{1/2})$ of the $D_1$ transition.

\end{document}